\documentclass{emulateapj} 
\usepackage{psfig}
\usepackage{apjfonts}
\usepackage{mathrsfs}
\shorttitle{New Metallicities of RRL Stars in $\omega$~Cen}
\shortauthors{Sollima et al.}

\begin{document}

\title{New Metallicities of RR Lyrae Stars in $\omega$~Centauri:
Evidence for a Non He-Enhanced Metal-Intermediate
Population\thanks{Based on FLAMES/GIRAFFE 
observations collected with the Very Large Telescope at the European Southern 
Observatory, Cerro Paranal, Chile, within the 
observing programs 74.B-0170(A).}}


\author{A. Sollima\altaffilmark{1}, J. Borissova\altaffilmark{2,3}, 
M. Catelan\altaffilmark{4}, H. A. Smith\altaffilmark{5}, D.
Minniti\altaffilmark{4},
C. Cacciari\altaffilmark{6} and F. R. Ferraro\altaffilmark{1}}

\altaffiltext{1}{Dipartimento di Astronomia, Universit\`a di Bologna, via Ranzani 1, I-40127 Bologna, Italy}
\altaffiltext{2}{European Southern Observatory, Avenida Alonso de C\'{o}rdova 3107, 763-0581 Vitacura, Santiago 19, Chile}
\altaffiltext{3}{Departamento de F\'isica y Meteorolog\'ia, Universidad de Valpara\'{\i}so,
 Ave. Gran Breta\~na 644, Playa Ancha, Casilla 53, Valpara\'iso, Chile}
\altaffiltext{4}{Pontificia Universidad Cat\'olica de Chile, Departamento de Astronom\'\i a y Astrof\'\i sica, Av. Vicu\~na Mackenna 4860, 
                 782-0436 Santiago, Chile}
\altaffiltext{5}{Department of Physics and Astronomy, Michigan State University, East Lansing, MI 48824}
\altaffiltext{6}{Osservatorio Astronomico di Bologna, via Ranzani 1, I-40127 Bologna, Italy}


\begin{abstract}
We present new spectroscopic metal abundances for
74 RR Lyrae stars in $\omega$~Cen obtained with FLAMES. The
well-known metallicity spread is visible among the RR Lyrae
variables. The metal-intermediate (MInt) RR Lyrae
stars (${\rm [Fe/H]} \sim -1.2$) are fainter than the bulk of
the dominant metal-poor population (${\rm [Fe/H]} \sim -1.7$), 
in good agreement with the corresponding zero-age horizontal branch
models with
cosmological helium abundance $Y = 0.246$. This result
conflicts with the hypothesis that the progenitors of the
MInt RR Lyrae stars correspond to the anomalous
blue main-sequence stars, which share a
similar metallicity but whose properties are currently
explained by assuming for them a large helium enhancement.
Therefore, in this scenario, the coexistence within the cluster 
of two different populations with similar
metallicities (${\rm [Fe/H]} \sim -1.2$) and different helium
abundances has to be considered. 
\end{abstract}


\section{Introduction}

The stellar system $\omega$~Centauri (NGC 5139) is the
most massive and luminous globular cluster of the Milky
Way, and is the only one showing a clear metallicity
spread. Extensive spectroscopic surveys performed on large
samples of giant stars (Norris et al. 1996;  
Suntzeff \& Kraft 1996) have found a multimodal
distribution of heavy elements. Recent photometric surveys
have revealed the presence of multiple sequences in the
color-magnitude diagram (CMD) of $\omega$~Cen. In
particular, a discrete structure of the red giant branch
(RGB) and subgiant branch (SGB) has been evinced (Lee
et al. 1999; Pancino et al. 2000; Rey et al. 2004; Ferraro
et al. 2004; Sollima et al. 2005a), indicating a complex
formation history. In addition to the dominant metal-poor
population (MP, ${\rm [Fe/H]} \sim -1.7$), three
metal-intermediate (MInt) components (spanning a range
of metallicity  $-1.3 < {\rm [Fe/H]} < -1.0$) and an
extreme metal-rich population (${\rm [Fe/H]} \sim -0.6$) have
been identified (Norris \& Da Costa 1995; Smith et al.
2000; Pancino et al. 2002; Vanture et al. 2002). 
In the faint part of the CMD, Anderson (2002) and Bedin et
al. (2004) discovered an additional blue main sequence
(bMS, comprising $\sim$ 30\%  of all MS stars) running
parallel to the dominant one. According to stellar models
with canonical chemical abundances, the location of the
observed bMS would suggest a lower metallicity. However,
Piotto et al. (2005), from a spectroscopic analysis of
34 stars belonging to both MS components, found that the
bMS stars are $\sim 0.3$~dex more metal-rich than the
dominant cluster population, and 
therefore might be associated with one of the MInt
components identified from the RGBs. The explanation
currently proposed to account for the anomalous position
of the bMS in the CMD is a significant helium enhancement 
(by $\Delta Y \sim 0.10 -0.15$) for this stellar population
(Norris 2004).

RR Lyrae stars (RRL), being representatives of the oldest populations, are 
important tracers of the formation history of $\omega$~Cen.
The first sign of peculiarities in the RRL variables of $\omega$~Cen was
found by Freeman \& Rodgers (1975), who observed a wide spread in
metallicity in a sample of 25 RRL. This result was later confirmed by 
Butler, Dickens, \& Epps (1978) and Gratton, Tornamb\`e, \& Ortolani (1986). 
More recently, Rey et al. (2000) derived metallicities for 131 RRL stars from 
the $hk$ index of the Ca$by$ photometric system. 

We present here new metallicities for 74 RRL stars of 
$\omega$~Cen, and discuss the implications of 
helium enhancement on the location in the CMD and the pulsational properties of 
the MInt group of RRL stars. 
%Our idea, in the present {\em Letter}, is to utilize the RRL stars,
%whose detailed properties, especially their luminosities and periods, are
%sensitive to the parameters involved in shaping the morphology of the
%horizontal branch, as a ``vertical diagnostic'' of horizontal branch
%morphology (Catelan, Sweigart, \& Borissova 1998). 
This can be 
useful in constraining the extent to which a spread in helium may or not
be present in the cluster.

\section{Data Analysis}

Observations of 74 RRL variables (38 RRab and 36 RRc) were
collected on February and March 2005 at the VLT/UT2 at ESO (Cerro
Paranal, Chile) with the multi-fiber spectrograph
FLAMES/MEDUSA.  We used the
high-resolution grating HR13 with a spectral coverage of
285~\AA\ 
(6120-6405~\AA) and a resolving power  $R \sim 22500$. 
For each star we obtained three exposures (3000~s + 
3600~s + 3600~s) secured in good seeing conditions ($FWHM <
0.8\arcsec$), reaching an average $S/N~\sim~40$ per pixel. The
one-dimensional spectra were extracted using the GIRAFFE
pipeline. Thirty fibers  were dedicated to sky observations in
each exposure. An average sky spectrum was obtained and
subtracted from the object spectra, by taking into account
the different fiber transmission. The spectra were then
continuum normalized with IRAF.

A set of 26 metal lines was selected  from the database of
Kurucz \& Bell (1995) in the spectral range covered by  our
spectra.  These lines were identified, whenever possible,
on each spectrum, and were fit with Gaussian functions,
whose centers provided the observed wavelength. The integral 
of the difference between the continuum and the line profile 
provided the equivalent width (EW) of each line.  The 
abundance analysis was performed using the latest version
of Kurucz (1979) model atmospheres\footnote{The effect of a helium overabundance
of $\Delta Y = 0.1$ on the model atmosphere is to
decrease the continuum opacity, thus enhancing the line strength by $\sim 5\%$
(corresponding to $\sim 0.03$~dex; B\"ohm-Vitense 1979). Given the small impact of
such an effect, we adopted model atmospheres with a standard helium abundance.} 
and the MOOG line analysis code (Sneden 1973) to compute LTE abundances from
individual EWs.  The abundance was derived from the model
that best reproduced the  observed EWs, for assumed values
of temperature, gravity and microturbulence velocity. The
errors on the abundances were calculated as the standard
deviation of the abundance relative to all the measured
lines.

Temperatures were derived from $V\!-\!K$ colors during the
phase interval between the beginning and the end of each
observation, using the $V$ light curves provided by Kaluzny
et al. (2004). The mean $K$ magnitude of each
star was derived from the infrared catalog by Sollima et al.
(2004) and corrected for the phase effect using the
template $K$ light curves by Jones et al. 
(1996).\footnote{Dereddened $(V\!-\!K)_0$ colors were
computed adopting the reddening and extinction coefficients
$E(\bv)=0.11$ (Lub 2002), $A_V=3.1~E(\bv)$ and
$A_K=0.38~E(\bv)$ (Savage \& Mathis 1979). Temperatures
were then derived using the
color-temperature conversions by Montegriffo et al. (1998).}
As an
independent check, we derived the temperature of each star
directly from  its spectrum by minimizing the trend in the
deduced abundances with the  excitation potential for the
observed Fe~{\sc i} and Fe~{\sc ii} lines.  The two
measurements agree with each other to within $\sim 100$~K 
for most of the RRL in our sample. 

The gravity parameter $\log g$ was then calculated from the universal gravitation 
law.\footnote{The gravity was calculated using the distance modulus $(m\!-\!M)_0=13.70$ 
(Bellazzini et al. 2004), the reddening and 
extinction coefficients listed above and the bolometric corrections by
Montegriffo et al. (1998).
The mass of the RRL stars was calculated from the stellar
pulsation theory using the relations of Marconi et al. (2003, their eq. 2).}
The microturbulence velocity was initially set as $\xi=2~{\rm km~s}^{-1}$, following 
Fokin \& Gillet (1997), and then adjusted within a range of about $1~{\rm km~s}^{-1}$ 
by minimizing the trend in the deduced abundances
with EWs for Fe~{\sc i} and Fe~{\sc ii} lines.

We performed the spectral analysis independently on each exposure adopting the 
extreme values reached by the parameters during the observation, and 
assumed the corresponding abundances as upper and lower limits. 
Spectra taken during maximum light were rejected because of non-LTE effects 
due to shock wave phenomena. 

\section{Results}

\begin{figure}
\plotone{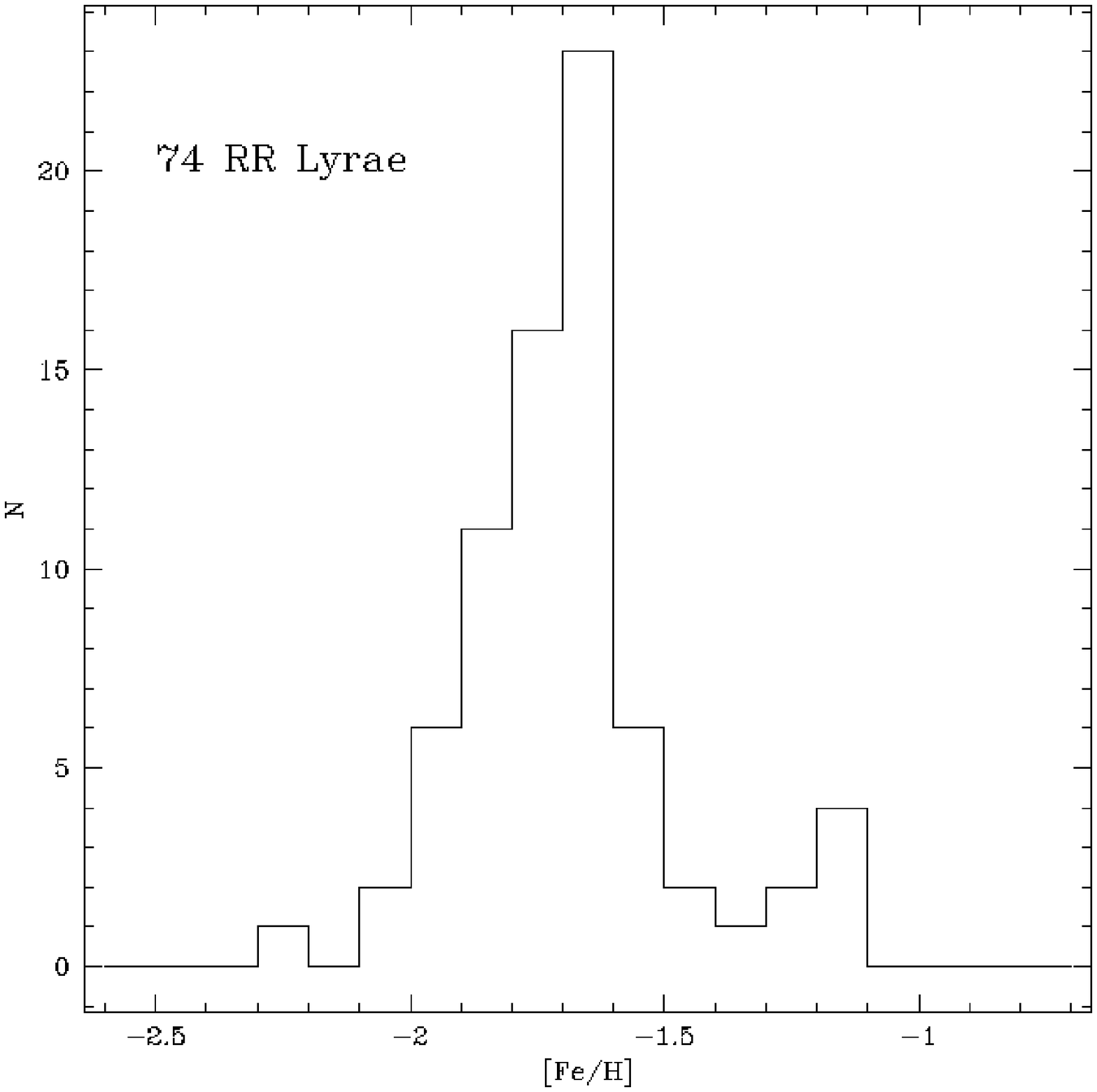}
\caption{Metallicity distribution for the 74 observed RRL stars.} 
\label{hist}
\end{figure}

The metallicities derived for the 74 observed RRL stars are listed in Table~1.
The metallicity distribution is shown in Fig.~\ref{hist}. The KMM test 
(Ashman, Bird, \& Zepf 1994) performed on our metallicity  distribution
rejects the single Gaussian model at the 99.9\% confidence level, supporting a bimodal distribution 
that peaks at ${\rm [Fe/H]}=-1.72$ and $-1.22$ (MP and MInt groups, respectively) with an
estimated common covariance 0.02.
A detailed comparison with the metallicities measured by Rey et al. (2000)
indicates a systematic offset 
$\Delta {\rm [Fe/H]} \mbox{(this work--Rey)} = -0.06$~dex, with a dispersion
of $\sim 0.3$~dex.
Since the accuracy of both sets of results is $\sim 0.3$~dex,  
the observed dispersion is consistent with the above uncertainties. 
%One of the strongest spectral features evident in the covered spectral range is
%the Ba II line at $\lambda~6141.7 \AA$. This feature allowed us to measure
%$[Ba/Fe]$ abundance ratios for most of the RRL stars of our sample. 
%Unfortunately, since the analysis is based only on a single Ba line, the obtained abundances 
%can be interpreted only in a relative sense. We observed that all the 10
%sub-luminous (i.e. metal-rich) RRL stars present an overabundance of
%$\Delta [Ba/Fe]~\sim~1.0~dex$ with respect to the bulk of metal-poor RRL.
%This result is in agreement with the results obtained on red giants by 
%Pancino et al. (2002).

\begin{figure}
\plotone{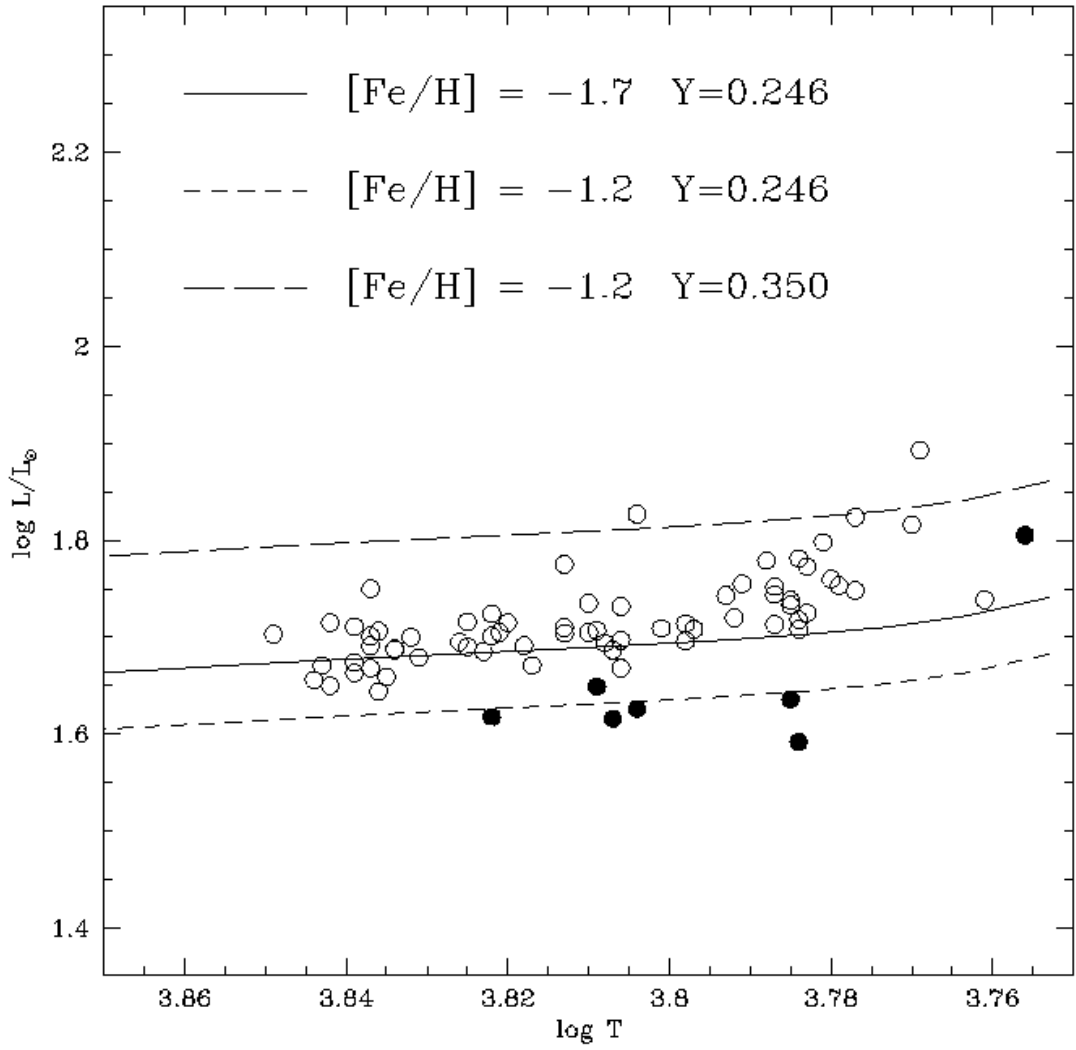}
\caption{HR diagram for the target RRL stars of $\omega$~Cen. Open circles mark stars 
with ${\rm [Fe/H]}<-1.35$, filled circles mark stars with ${\rm [Fe/H]}>-1.35$. 
Different lines indicate the ZAHB luminosity for the indicated metallicity and 
helium abundance combinations. } 
\label{hr}
\end{figure}

The location in the HR diagram of the RRL stars in our
sample\footnote{We transformed the $M_K$ magnitudes and
$(V\!-\!K)_0$ colors into luminosities and temperatures using
a distance modulus $(m\!-\!M)_0=13.70$  (Bellazzini et al.
2004), as well as color transformations and $K$-band bolometric
corrections by Montegriffo et al. (1998). The colors were
corrected for the non-static  atmosphere using the
correction by Bono, Caputo, \& Stellingwerf (1995).} is shown in 
Fig.~\ref{hr}. As can be seen, the dominant MP RRL population
(${\rm [Fe/H]} \sim -1.7$) has a mean luminosity
$\langle \log(L/L_{\odot})\rangle \sim 1.69$, whereas MInt RRL 
(${\rm [Fe/H]} \sim -1.2$) are systematically fainter, with 
$\langle \log(L/L_{\odot})\rangle \sim 1.62$. This evidence is in agreement
with the results by Rey et al. (2000). For comparison, we
show three zero-age horizontal branch (ZAHB) 
models\footnote{The ZAHBs were calculated
using the FRANEC code (Straniero, Chieffi, \& Limongi
1997).  The location in the CMD was interpolated from the
tracks of stars with  masses ranging from 0.52 to 0.80~$M_\odot$.} 
with ${\rm [Fe/H]}=-1.7$ and $Y=0.246$,  ${\rm [Fe/H]}=-1.2$
and $Y=0.246$, and ${\rm [Fe/H]}=-1.2$ and $Y=0.35$. The ZAHB models
with $Y=0.246$ fit nicely the lower envelopes of the 
respective MP and MInt RRL stars. Conversely, the
helium-enhanced ($Y=0.35$) MInt model  is far too bright and
does not match the observed MInt RRL stars.  

\begin{figure}
\plotone{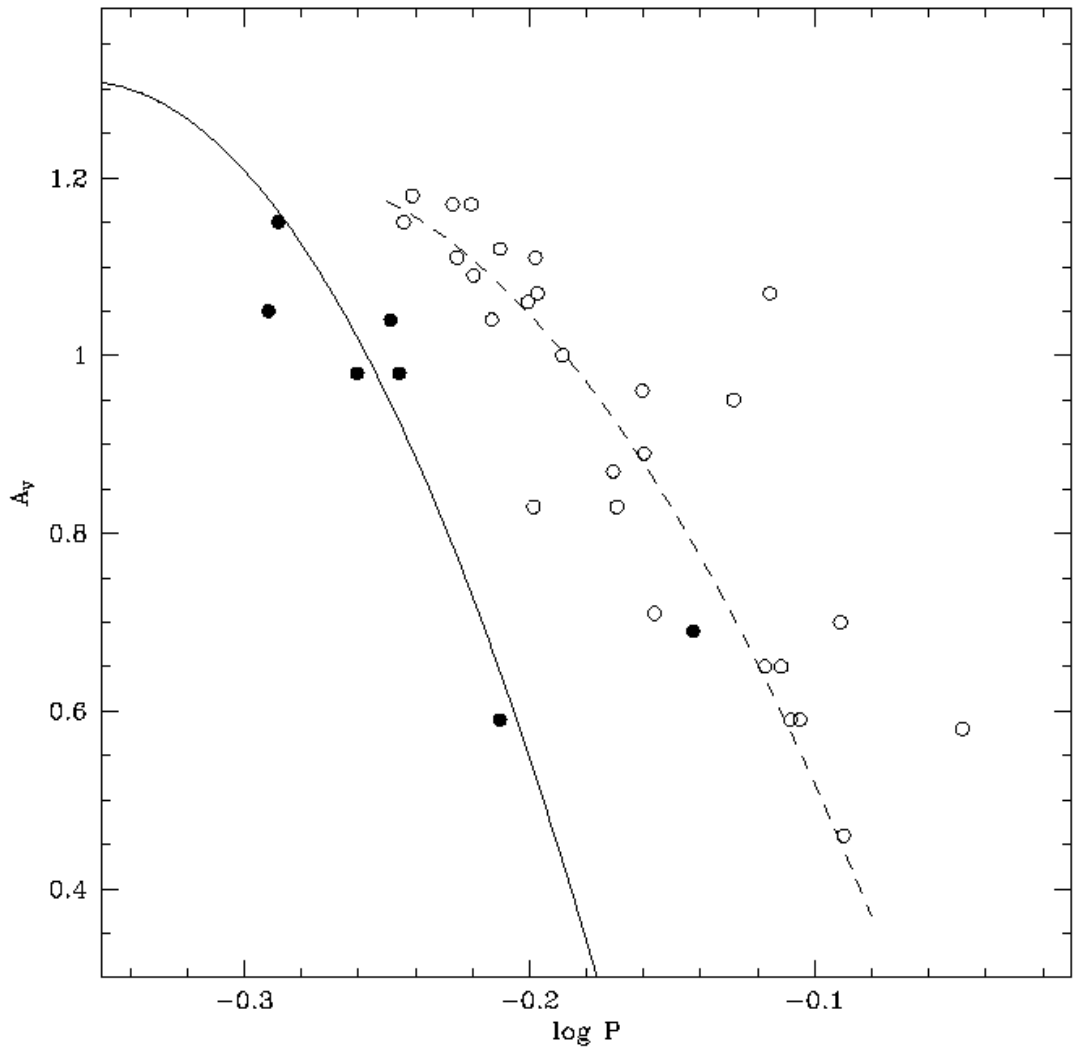}
\caption{The period-$V$ amplitude relation for the two metallicity groups of RRab
Lyrae in $\omega$~Cen. The solid line indicates the mean locus of the M3 variables, 
while the dashed line indicates the mean locus of the M92 variables.} 
\label{Oosth}
\end{figure}

Fig.~\ref{Oosth} shows the location of the RRab Lyrae in our sample in the
period-amplitude diagram. For comparison, the mean locus of the M3 (Oosterhoff I
type; Cacciari et al. 2005) and M92 (Oosterhoff II type) RRab Lyrae
variables are overplotted in this figure.
As can be noted, the stars of the MInt and MP samples occupy 
different regions of the diagram: while MInt stars do not show significant deviations from 
the M3\footnote{Note that M3 is a good reference cluster 
for the MInt RRL population in spite of its lower metal abundance, 
because metallicity does not have a significant effect on the period-amplitude 
distribution in globular clusters belonging to the same Oosterhoff group
(Cacciari et al. 2005).} period-amplitude relation, MP stars have 
longer periods and fall nicely on the M92 period-amplitude relation. 
. In this respect, the effect of a He enhancement on 
the MInt RRL would be to produce longer periods, as a consequence of the 
higher luminosities involved (Catelan 1996).

\begin{figure}
\plotone{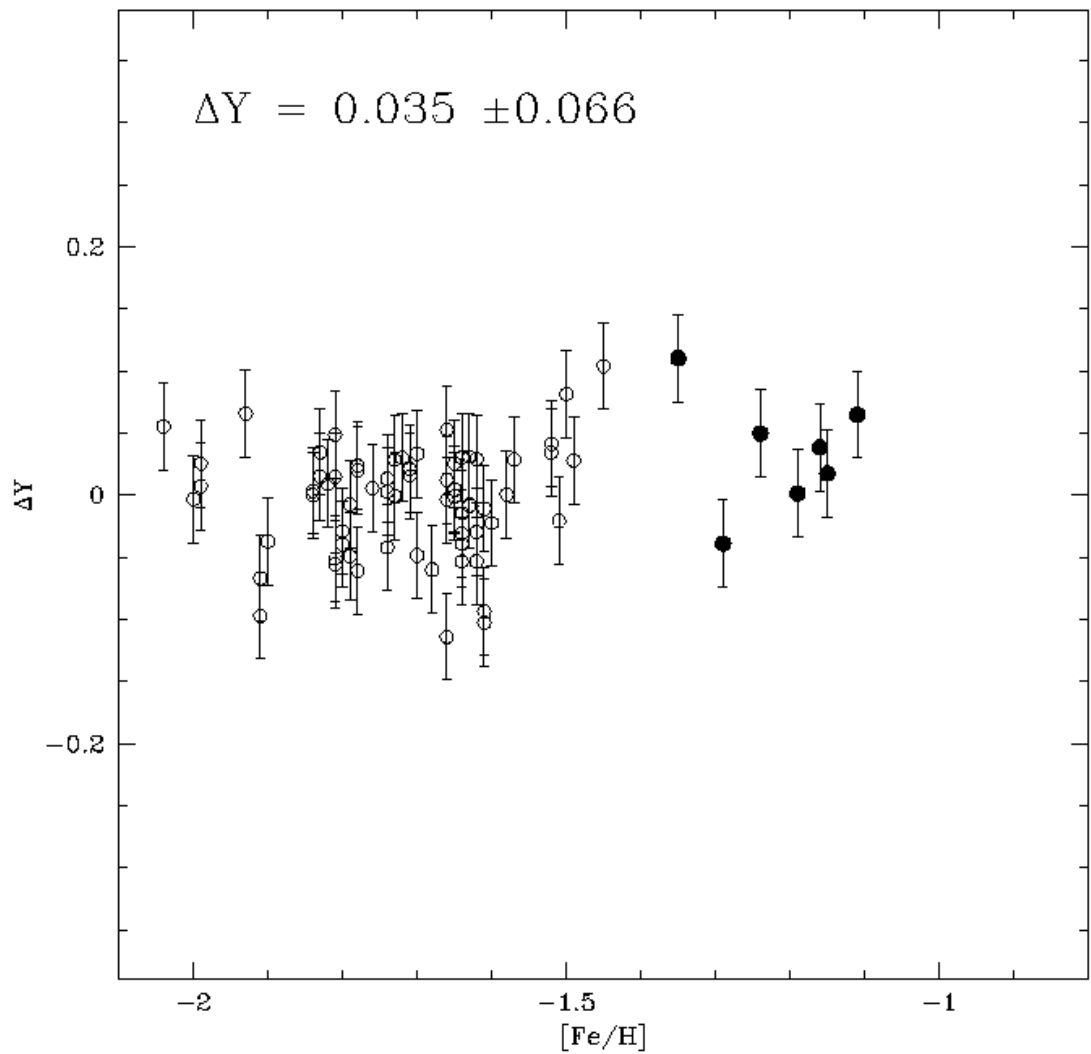}
\caption{Relative He abundances computed via the $A$ parameter for the observed 
RRL stars as a function of metallicity. Symbols are as in Fig.~2} 
\label{helium}
\end{figure}

An independent check on the helium abundance can be made using 
the mass-luminosity parameter $A$ (Caputo, Cayrel, \& Cayrel
de Strobel 1983; Sandquist 2000).
%\begin{eqnarray*}
%  A & = & 0.81~\log\left(\frac{M}{M_\odot}\right) - \log\left(\frac{L}{L_\odot}\right) \\ 
%    & = & 13.353-1.19~\log\,P+4.058~\log\,T_{eff}.  
%\end{eqnarray*}
%\noindent 
This parameter
depends on the helium abundance in the sense that, at a fixed 
mass and for a given temperature, increasing the helium abundance 
increases the luminosity, and thereby the period, of an RRL  
star. Although $A$
has a relatively low sensitivity to the helium abundance,
statistical errors are generally small ($\sigma_A \sim
0.05$), and this method may provide some useful
indications in regard to (at least) relative helium abundances. 
We derived $A$ values for the 74 RRL stars in our sample using eq. 8 of 
Cacciari et al. (2005), and
converted these values into $Y$ differences with respect to
the mean $Y$ of the MP stars assuming
$\delta A/\delta Y = 1.4$ (Sandage 1990; Bencivenni et al.
1991). Figure~\ref{helium} shows the obtained values of $\Delta Y$ 
as a function of metallicity. As can be seen, there is no evidence 
of a systematic trend with metallicity to within the errors 
($\sigma_Y = 0.07$). 

\section{Discussion}

The most important result of this analysis is the
confirmation of the existence of a metal-intermediate
(${\rm [Fe/H]}\sim-1.2$) RRL population in $\omega$~Cen with 
luminosity and pulsational properties that are {\em
incompatible with any  significant helium overabundance}.
This evidence adds new complexity to the scenario proposed
to explain the origin of the bMS. In fact, while
spectroscopic analyses suggest that the bMS (Piotto
et al. 2005) and the MInt RRL share a similar
metallicity (${\rm [Fe/H]} \sim -1.2$), the results presented in
this {\em Letter} demonstrate that they cannot share the same
helium abundance. On the other hand, a helium-rich  
population is not expected to produce a sizeable RRL component 
(Lee et al. 2005), unless a fairly large age difference with 
respect to the dominant population of $\omega$~Cen ($\Delta
t \gtrsim 4$~Gyr, in the sense that the He-enriched population 
should be younger) is assumed, in contrast with the results by 
Hilker et al. (2004) and Sollima et al. (2005b), who both 
found a rather small ($\lesssim 2$~Gyr) age difference between 
these two populations of $\omega$~Cen.

Therefore, following these considerations, two populations with
similar metallicities but very
different helium abundances seem to coexist within the cluster.
This possibility is still compatible with the
complexity of the  population mix observed at the SGB level
(see Ferraro et al. 2004; Sollima et al. 2005b), and sets
severe constraints on the helium enrichment mechanism  
and its timescale. 

As a matter of fact, in the framework of the self-enrichment 
scenario, the huge 
helium enrichment required to explain the
anomalous position of the bMS in the CMD has to occur 
after the entire MP and (at least part of) 
the MInt population (comprising $\sim 70\%$ of the whole cluster population) 
were already formed without any significant helium enhancement.
Then, the MInt He-rich population has to be formed from a gas enriched in 
helium by $\Delta Y\sim 0.10-0.15$ maintaining its metal
abundance practically unchanged.
Under this assumption, if the sources of the helium enrichment are
intermediate-mass stars with $M \sim 3-6 \, M_{\odot}$ 
(as suggested by D'Antona \& Caloi 2004), then 
the entire chemical enrichment of the system has to occur  
over a time-scale of $\sim 50-300$~Myr. Moreover, a
very efficient mechanism is required in order to homogenize and
efficiently reuse all the ``enriched'' material ejected by the
previous  generation of polluting stars (Norris 2004). 
In this respect, further support to this scenario could come from 
new evidence suggesting that a similar helium self-enrichment
may have occurred also in the metal-homogeneous  
globular cluster NGC~2808 (Lee et al. 2005; D'Antona
et al. 2005; see also Catelan 2006 for a recent, critical 
discussion). Also in this case, no He-rich RRL have been found in the cluster 
(Corwin et al. 2004; D'Antona et al. 2005)

Alternatively, scenarios outside the ``pure self-enrichment'' one 
may also be considered; in particular, one possibility could be that the MInt 
populations formed in a different environment, thus not partaking the 
chemical enrichment process of $\omega$~Cen. In this 
case, a complex interplay of chemical and dynamical evolution has to be 
taken into account, including gas exchange with the Milky Way and/or minor 
mergers, within a framework such as the binary cluster mergers scenario 
(Minniti et al. 2004).

Alternative scenarios which try to reproduce the anomalous location of the bMS 
without invoking a helium overabundance 
are not much simpler and present even more
obvious inconsistencies with observations. For example, if we assume that
MInt RRL and bMS belong to the same population (with the
same cosmological helium abundance), then the anomalous position of the bMS
can only be explained by assuming
a different (longer) distance modulus [$\Delta
(m\!-\!M)_0\sim~0.4$~mag, corresponding to $\sim 1.1$~kpc behind
the main body of the cluster]. This would be 
inconsistent with  the MInt RRL luminosity shown in Fig. 2, 
which suggests that MInt RRL are at the same distance as the
MP population. Moreover, Piotto et al. (2005) found that bMS stars
show the same radial velocity distribution of the MP group,
implying that the bMS and the main body of $\omega$~Cen are
at least dynamically bound to each other.

The results presented in this analysis add new questions in
the interpretation of the complex metal enrichment history
of $\omega$~Cen. However, this additional piece of 
information is important to define a scenario able to explain
the inhomogeneous chemical enrichment of the stellar
populations of this system.  

\acknowledgments

This research was supported by the Ministero dell'Istruzione Universit\`a e Ricerca.  
We thank the anonymous referee for his helpful comments and suggestions.
Support for M.C. was provided by Proyecto FONDECYT Regular No.~1030954.
D.M. is grateful for the support 
by the FONDAP Center for Astrophysics 15010003, by the European Southern Observatory
Visitor programme, and by a Fellowship from the John
Simon Guggenheim Foundation. HAS thanks the NSF for support under
grant AST 0205813.

\clearpage

\clearpage

\clearpage

\clearpage

\clearpage

\begin{deluxetable}{lccccccccc}
\tablewidth{0pt}
\tablecaption{Metallicities and properties of the target RRL}
\tablehead{
\colhead{OGLE ID}              & \colhead{[Fe/H]}     &
\colhead{$\sigma_{\rm [Fe/H]}$}    & \colhead{$\langle V\rangle$}      &
\colhead{$\langle K \rangle$}                & \colhead{$\log\langle T\rangle$} &
\colhead{$\log\langle L/L_{\odot}\rangle$} & RRL type              &
\colhead{$A_{V}$}              & $P$(d)\\}
\startdata
 5   & -1.24 &  0.11 &  14.745 & 13.295 & 3.804 & 1.626 & 1 &  1.15 & 0.515\\ 
 11  & -1.61 &  0.22 &  14.534 & 13.158 & 3.813 & 1.704 & 1 &  0.65 & 0.565\\ 
 15  & -1.68 &  0.18 &  14.397 & 12.765 & 3.784 & 1.781 & 1 &  0.70 & 0.811\\ 
 16  & -1.65 &  0.46 &  14.562 & 13.409 & 3.839 & 1.674 & 0 &  0.49 & 0.330\\ 
 20  & -1.52 &  0.34 &  14.579 & 12.953 & 3.784 & 1.708 & 1 &  1.12 & 0.616\\
\enddata
\tablecomments{The complete version of this table is in the electronic edition of
the Journal.  The printed edition contains only a sample.}
\end{deluxetable}

\end{document}